\documentclass[10pt]{article}
\usepackage{epsf} 
\usepackage{amsfonts}
\begin{document}
\newcommand{\be}{\begin{equation}}
\newcommand{\ee}{\end{equation}}
\newcommand{\kk}{{\bf k}}
\newcommand{\kkt}{\kk^\perp}
\newcommand{\p}{{\bf p}}
\newcommand{\q}{{\bf q}}
\newcommand{\qt}{\q^\perp}
\newcommand{\X}{({\bf x})}
\newcommand{\Y}{({\bf y})}
\newcommand{\x}{{\bf x}}
\newcommand{\y}{{\bf y}}
\newcommand{\U}{{\bf u}}
\newcommand{\w}{{\bf \omega}}
\newcommand{\D}{{\bf \nabla}}
\newcommand{\W}{\omega_\kk}
\newcommand{\za}{\alpha}
\newcommand{\zb}{\beta}
\newcommand{\zs}{\sigma}
\newcommand{\zt}{\tau}
\newcommand{\zN}{I\hskip-3.4pt N}
\newcommand{\zR}{I\hskip-3.4pt R}
\newcommand{\zw}{\omega}
\newcommand{\zC}{{\mathbb C}}
\newcommand{\OM}{({\bf ***\ldots***})}
\newcommand{\EM}{({\bf $\leftarrow$***})}
\newcommand{\BM}{({\bf ***$\rightarrow$})}

\title{Fluctuations in 2D reversibly-damped turbulence}
\author{Lamberto Rondoni \\
Dipartimento di Matematica, Politecnico di Torino, Italy\\
\tt{rondoni@calvino.polito.it}
\and Enrico Segre \\
Dipartimento di Ingegneria Aeronautica e Spaziale, Politecnico di Torino, 
Italy\\
\tt{segre@athena.polito.it}}
\date{\today}
\maketitle

\begin{abstract}
Gallavotti proposed an equivalence principle in
hydrodynamics, which states that forced-damped
fluids can be equally well represented by means of the Navier-Stokes
equations and by means of time reversible dynamical systems called
GNS. In the GNS systems, the usual viscosity is
replaced by a state-dependent dissipation term which fixes one global
quantity. The principle states that the mean values of properly
chosen observables are the same for both representations of the
fluid. In the same paper, the chaotic hypothesis of Gallavotti and
Cohen is applied to hydrodynamics, leading to the conjecture that
entropy fluctuations in the GNS system verify a relation first
observed in nonequilibrium molecular dynamics. We tested these ideas
in the case of two-dimensional fluids. We examined the fluctuations of
global quadratic quantities in the statistically stationary state of
a) the Navier-Stokes  equations; b) the GNS
equations. Our results are consistent  with the validity of the
fluctuation relation, and of the equivalence principle,  indicating
possible extensions thereof. Moreover, in these results the
difference between the Gallavotti-Cohen fluctuation theorem and the
Evans-Searles identity is evident.
\end{abstract}

\begin{center} 
PACS1998 numbers: 47.27.Gs, 05.40.+j, 05.70.Ln
             
AMS1991 numbers: 82C05, 76F20 
\end{center}

\section{Introduction}
The fundamental laws governing the behaviour of fluids are known, and
universally believed to be correctly represented by the Navier-Stokes
(NS) equations.  However, a clear cut connection between microscopic
and macroscopic scales, which definitely justifies such a belief, has
still to be made, particularly for systems subjected to
nonconservative external fields \cite{Sp91}. If one considers this
kind of problems as merely technical, and accepts the NS equations as
a valid tool for the description of a fluid's dynamics, serious
difficulties are met nevertheless in the study of the mathematical
properties of such equations. For instance, the problem of global
existence and regularity of solutions of the NS equations is far from
being solved in as general terms as desired. This fact is not of
exclusively mathematical interest, because it casts doubts on the
effectiveness of the approximation methods devised to extract
information from the NS equations. Therefore, what is reasonable  and
presently feasible is to study simplified problems, and to construct
theories connecting such problems with the full NS equations or with
the  real dynamics of the fluids at hand. In Ref.\cite{SJ}, this is
stated as: {\it ``It has been recognized that a realistic goal for a
statistical theory of turbulence is to determine the equations
governing the dynamics of some reduced set of modes which allow
calculation of fundamental quantities of the flow''}.

The situation is similar, to some extent, to that of molecular
dynamics, in nonequilibrium statistical mechanics. There, some
progress took place  when infinite reservoirs, or driving boundary
conditions were replaced by  artificial constraints imposed on the
bulk dynamics of $N$-particle systems,  with $N < \infty$ (see e.g.\
\cite{No84,Ho85,EM,GalCoh95}).  This way direct numerical simulations
of the particle models become feasible, and dynamical systems  theory
leads to theoretical predictions \cite{GalCoh95} which can be tested
in numerical simulations or real experiments. The trade-off of this
approach  is that the relevant dynamical  equations do not seem to be
fully justified on physical grounds: no known fundamental force acts
on the particles of the system in such a way as to implement the
desired constraints. Nonetheless, the  results obtained this way are
in excellent agreement with experience \cite{EM}, and several
arguments have been developed to explain why this should be the case.
Among such arguments, we deem more convincing those invoking a kind of
equivalence of ensembles, which is a well known concept in equilibrium
statistical mechanics, but rather new in nonequilibrium statistical
mechanics \cite{Gal95,Gal97}. If verified, the  equivalence of
ensembles guarantees that different microscopic dynamics result  in
the same macroscopic behaviour, thus justifying the use of one kind of
dynamics or another, depending on which one is more natural to study a
given physical problem.\footnote{Consider, for instance,  the success
of the idealized dynamics of lattice gas cellular automata in
describing quite complex hydrodynamic situations \cite{RZ}.}

Similarly, in Ref.\cite{Gal97}, Gallavotti argues that different probability
distributions characterizing the steady state properties of a fluid should 
yield the same values for (some, at least) macroscopic quantities, if such 
distributions are obtained from different microscopic models of
the same fluid.
In particular, inspired in part by the mentioned developments in 
statistical mechanics \cite{GalCoh95} and by Ref.\cite{SJ},
Gallavotti conjectures one kind of equivalence between the NS equations 
and special time reversal invariant equations which he calls GNS,
where the G stands for Gaussian constraint.

The paper \cite{SJ} introduces a constrained Euler system, which
encompasses a portion of the degrees of freedom of the NS equations
sufficient to obtain some of the fundamental statistical 
properties of the fluid. The explicit form of the constrained
equations is derived from the observation that in stationary isotropic
turbulence, the mean energy in a narrow wave-number shell is nearly
constant in time. These equations are similar to those
used in nonequilibrium molecular dynamics, for driven particle
systems subjected to a ``gaussian thermostat''
\cite{No84,Ho85,EM}. Therefore, the properties of such  particle
systems should be observed to some degree in the constrained Euler
system of \cite{SJ} and, perhaps, in still more general settings. This
led  Gallavotti \cite{Gal97} to revive and put under a new light
Ruelle's principle  for hydrodynamics \cite{Ru78}. This principle had
been previously extended to  nonequilibrium statistical mechanincs in
\cite{GalCoh95}, where the first  proof of the Gallavotti-Cohen
Fluctuation Theorem (GCFT) was obtained, under  the assumptions of the
Chaotic Hypothesis (CH) quoted below in Section 2.

Tests of the validity of the equivalence conjecture (EC) and of 
the CH include a Rayleigh-Benard convection experiment by S.\ Ciliberto and
C.\ Laroche \cite{CL}, and numerical simualtions of the GOY shell models 
by L. Biferale, D. Pierotti and A. Vulpiani \cite{BiPiVu97}. The experiment 
of \cite{CL} is consistent with the validity of a fluctuation relation 
\cite{Gal97} similar to that of the GCFT; while the simualtions of 
\cite{BiPiVu97} evidenced a kind of equivalence of different hydrodynamic 
models: the equivalence of energy cascades. A posteriori, the 
work by She and Jackson \cite{SJ} can also be taken as a verification of the 
validity of the EC, although apparently not motivated by a general theory.  
Among other relevant works, notable are Refs.\cite{BoGaGa96,LLP,BoChLe98},
although they do not concern hydrodynamic equations.

In the present paper, we test both the validity of the  mentioned
fluctuation relation and of the EC for the NS and GNS
equations. Because lengthy calculations are needed, we consider
two-dimensional systems rather than three-dimensional ones. Our
results confirm, and actually indicate possible extensions of both the
EC and the fluctuation relation of Ref.\cite{Gal97}, consistently with
Gallavotti's predictions for the slope of such (linear)
relation. Furthemrore, as explained in Section 5, these results
provide an example in which the difference between the GCFT and an
identity previously obtained by Evans and Searles \cite{ES94} is
evident.

\section{Reversible damping and the equivalence conjecture}
The CH has been introduced \cite{GalCoh95} in the study of 
$N$-particle systems subjected to nonconservative external forces 
and to ``gaussian'' constraints \cite{No84,Ho85,EM}, like dynamical 
systems of the form
\be
  \dot{\bf q}_i = {\bf p}_i/m_i ~, \quad
  \dot{\bf p}_i = {\bf F}_i^{\mbox{int}} + c_i {\bf F}^{\mbox{ext}} - 
  \za {\bf p}_i ~, \quad \mbox{for} \quad i = 1, ..., N  
\label{eqsmot}
\ee
defined on a $(2dN-1)$-dimensional manifold $\Omega \subset \zR^{2dN}$, $d$ being
the dimension of the physical space. Here, $({\bf q}_i,{\bf p}_i)$ is the usual
notation for position and momentum of particle $i$, ${\bf F}_i^{\mbox{int}}$
represents the action of the other $N-1$ particles on particle $i$, $c_i$ is a
charge coupling particle $i$ to the external field ${\bf F}^{\mbox{ext}}$, and
$\za {\bf p}_i$ is a dissipation term, which allows the system to reach a
stationary state. In particular, the choice
\be
  \za = \frac{1}{2K} {\bf F}^{\mbox{ext}} \cdot \sum_{j=1}^N 
  c_j {\bf p}_j ~,
\ee
where $K = \sum_{i=1}^N {\bf p}_i^2/ 2 m_i$ 
is the kinetic energy of the system, implies that the internal energy of the 
system $E = K + \Phi$, $\Phi$ being the internal potential energy, is a
constant of the motion. 
We denote by $V_t : \Omega \rightarrow \Omega$, $t \in \zR$, the flow, so that 
$t\mapsto V_t \gamma$ represents a solution of Eqs.(\ref{eqsmot}) with initial
condition $\gamma$. The dynamical system $(V_t,\Omega)$ is but an example from 
a wide class of systems which share remarkable properties \cite{EM,GalCoh95}. 
For example, such systems are time reversal invariant, although dissipative.
This means that the
involution $i : \{{\bf p}_j\}_{j=1}^N \mapsto \{-{\bf p}_j\}_{j=1}^N$
anticommutes with the time evolution: $i V_t = V_{-t} i$ (reversibility); but
phase space volumes contract on average (dissipativity), resulting in a
multi-fractal stationary state \cite{HoHoPo87}. Such systems have been 
studied in detail, under the assumption that the following holds
\cite{Gal97}:

\vskip 5pt
\noindent
{\bf Chaotic Hypothesis (CH).} {\it A chaotic many-particle system or {\em 
fluid} in a stationary state can be regarded, for the purpose of computing
macroscopic properties, as a smooth dynamical system with a transitive
Axiom-A global attractor. In reversible systems it can be regarded, for
the same purpose, as a smooth transitive Anosov system.}

\vskip 5pt
\noindent
This approach led to interesting results for the macroscopic properties of given
systems, directly from their microscopic dynamics. Among these are proofs of 
the positivity of transport coefficients \cite{LlNiRoMo95,Ruelle96}, of the 
validity of Onsager relations\footnote{The microscopic 
models considered in these studies constitute a wide 
class with some peculiar feature, such as the unusual 
dissipation terms. Therefore, it is not immediately clear that 
locutions borrowed from physics --such as ``Onsager relations''-- can 
be attributed the same phenomenological meaning they usually have. For 
instance, a minimal prerequisite for the models to be physically relevant 
is that their number of particles be large \cite{CoRo98}, while this 
is not assumed in \cite{LlNiRoMo95} -- \cite{RoCo}.} \cite{GaPRL,RoCo}, and 
of the fluctuation relation of \cite{GalCoh95}, whose hydrodynamic version 
Eq.(\ref{FluTh}) is given below.

Following Gallavotti's ideas \cite{Gal97}, we now consider 
the NS equations for a newtonian incompressible fluid:
\be
\dot{\bf u} + ({\bf u} \cdot \D) {\bf u} = 
- \frac{1}{\rho} \D p + {\bf g} + \nu_1 \Delta {\bf u}
~, \quad \D \cdot {\bf u} = 0 ~.
\ee
Here, {\bf u} is the velocity field, $\rho$ is the fluid density,
$p$ the pressure, {\bf g} is a constant forcing term and $\nu_1$ is the
constant viscosity. The curl of this equation gives
\be
\dot{\bf \omega} +  ({\bf u} \cdot \D){\bf \omega} =
 ({\bf \omega} \cdot \D) {\bf u}+
 {\bf f} + \nu_1 \Delta {\bf \omega}   \label{Helm}
\ee
in which ${\bf \omega} = \D \times {\bf u}$ is called vorticity
and ${\bf f}=\D \times{\bf g}$ represents the forcing term.
If we replace $\nu_1$ by 
\be
\beta_Q({\bf u},{\bf \omega},{\bf f}) = \frac{\int \left[
\w \cdot {\bf f}   + {\bf \omega} \cdot
({\bf \omega} \cdot \D) {\bf u} \right] ~ d {\bf x}}{
\int \left(\D \times {\bf \omega} \right)^2 d {\bf x}}\,, \label{betau}
\ee
we obtain a system whose total enstrophy $Q = \int \omega^2 d {\bf x}$
is a constant of motion, and which is time reversal invariant in the 
sense Eqs.(\ref{eqsmot}) are. Similarly to Ref.\cite{Gal97}, we refer to 
Eq.(\ref{Helm}) with $\beta_Q$ in place of $\nu_1$, as to the GNS equations, 
i.e. the NS equations with a Gaussian constraint. 
In two spatial dimensions, 
the vortex-stretching term $(\w\cdot\D)\U$ vanishes; hence 
only a dynamical equation for the third component of 
the vector $\w$ is necessary. Denoting this component also 
by $\omega$, the vorticity equation reduces to
\be
  \dot{\omega}= -({\bf u} \cdot \D){\omega} +
   f + \alpha \Delta {\omega} \,.                \label{Helm2d} 
\ee
where $\alpha$ stands either for $\nu_1$ or for $\beta_Q$, depending on the
case. We impose doubly-periodic boundary conditions on
Eq.(\ref{Helm2d}), and rescale the size of the system to $2\pi$, making
natural a Fourier expansion for $\omega$. Such expansion,
truncated for numerical implementation, yields the approximation
\be
  \omega(\x) \approx \sum_{k_x=-N}^N\sum_{k_y=-N}^N e^{i\kk\cdot\x}\W \,,
  \quad  \quad \kk = (k_x,k_y) ~, \quad N \in \zN ~.
  \label{cutoff}
\ee
Substituting Eq.(\ref{cutoff}) in
Eq.(\ref{Helm2d}), we get the equations for the Fourier modes $\W$. 
We write such equations in the more general form:
\be
\dot{\w}_\kk = r_\kk
      + f_\kk - \alpha \kk^{2l} \W   \,, \quad k_x,k_y=-N,...,N ~, \quad
      l = 0,1,2, ... \label{NSk}
\ee
where each $l$ accounts for the dissipation given
by a power of the Laplace operator applied to $\omega$. This way, 
we get Ekman damping for $l=0$  ($\alpha=\nu_0$), normal viscosity for $l=1$ 
($\alpha=\nu_1$), and a different hyperviscosity $\nu_l$ for any $l>1$, 
while only $l=1$ was considered in \cite{Gal97}. 
The quantity $r_\kk$ is the nonlinear interaction
term, which reads
\be
r_\kk=\sum_\p
     \sum_\q{{\qt\cdot\p}\over{\q^2}} 
       \omega_\q \omega_{\p}\delta_{\kk,\p+\q} \label{nltk} ~;     
        \qquad \qt=(q_y,-q_x) ~. 
\ee
Equations (\ref{NSk}) determine the dynamics of the
$(2N+1)^2$ complex modes $\W$, in a phase space $\Omega \subset \zC^{(2N+1)^2}$
whose dimension is only $2N(N+1)$, because of the reality condition
$\W=\omega^*_{-\kk}$ and of the absence of the $\kk=(0,0)$ mode.

Differently from \cite{Gal97}, we consider all the quadratic global 
quantities, such as the energy $E=\int \omega \Delta^{-1} \omega d\x$, the 
enstrophy $Q=\int \omega^{2} d\x$, the palinstrophy $P=\int (\D\omega)^{2} d\x$, 
the ``hyperpalinstrophy" $H=\int (\Delta\omega)^{2} d\x$, etc, instead of
considering $E$ and $Q$ only. 
Using the spectral notation, we can write any one of them as 
\be
   Q_m = \sum_\kk \kk^{2m} \W^*\W \, ,
\ee
where $Q_{-1}=E$, $Q_0=Q$, $Q_1=P$, $Q_2=H$  etc. These quantities satisfy the
evolution equations
\be
  \dot{Q}_m=2 F_m  - 2\alpha Q_{l+m} +2 R_{m,0}~, \quad
  m = -1,0,1,2,...
\ee
where we have defined for convenience of notation
\begin{eqnarray}
     F_m &=& \sum_\kk \kk^{2m}\W^*f_\kk\, , \nonumber \\
     R_{m,l}&=&  \sum_\kk \sum_\p \sum_\q \w_\kk\, \w_\p\, \w_\q\,
\delta_{\kk+\p+\q,0}\,
              (\kk^{2l}+\p^{2l}+\q^{2l}) \frac{\qt\cdot\p}{\q^2}\kk^{2m}
          \label{Rmlkpq}
\end{eqnarray}
and $l$ depends on the kind of viscosity 
under consideration. The term $R_{m,0}$ vanishes for $m=-1$ and $m=0$, because
of the symmetries of the summands: this corresponds to the conservation
of energy and enstrophy in inviscid two-dimensional flow. Now, if we take
\be
 \alpha = \beta_{l,m}= {{F_m+R_{m,0}}\over{Q_{l+m}}} \,,
                      \label{revdissk}
\ee
instead of the constant viscosity coefficient $\nu_l$ in Eq.(\ref{NSk}),
$Q_m$ becomes a constant of motion. In this case, we refer to Eqs.(\ref{NSk}) 
as to the ``cut-off GNS equations''.
\footnote{Equations (\ref{NSk}), with $\za = \beta_{l,m}$,
for an arbitrary choice of $l,m$, cannot be interpreted in terms of Gauss'
principle, while that interpretation was allowed in the cases of 
Ref.\cite{Gal97}.}
An important quantity is $\sigma_\alpha$, the negative of the divergence 
in phase space of the right hand side of Eqs.(\ref{NSk}),
\be
\zs_\za = -\sum_\kk{\delta{\dot{\omega}_\kk}\over{\delta\omega_\kk}} =
   \alpha \sum_\kk \kk^{2l} + \sum_\kk \kk^{2l} \W \frac{\delta \alpha}
   {\delta\omega_\kk}\, ,
\ee
which may be identified \cite{GaPRL,CoRo98} with the entropy production rate 
of the system. Then, for $\za = \nu_l$
we get
\be
  \sigma_l^{NS} =  \nu_l \sum_{k_x=-N}^N\sum_{k_y=-N}^N \kk^{2l} -0^{2l} 
  \label{signu}
\ee
while for $\za = \zb_{l,m}$ (the GNS case) we have
\be
  \sigma_{l,m}^{GNS} =  \beta_{l,m} 
  \left(\frac{\sigma_\nu}{\nu_l}-2 \frac{Q_{2l+m}}{Q_{l+m}}\right) +  
  \frac{F_{m+l} + R_{m,l}}{Q_{l+m}} ~.
  \label{sigmarev}
\ee
The sums in Eq.(\ref{signu}) evaluate to
\begin{eqnarray}
(2N+1)^2 -1  \qquad  &\mathrm{for}\ & l=0\ \mathrm{(Ekman\ damping),} 
\nonumber \\
{{2}\over{3}}(2N+1)^2 (N+1) N  
      \qquad &\mathrm{for}\ & l=1\ \mathrm{(normal\ viscosity),} \\
{{2}\over{45}}(14N^2+14N-3)(2N+1)^2 (N+1) N 
    \qquad  &\mathrm{for}\ & l=2\ \mathrm{(hyperviscosity),\ etc.}  \nonumber 
\end{eqnarray}
Clearly, $\sigma_l^{NS}$ is a constant, while $\sigma_{l,m}^{GNS}$ depends on 
the vorticity field, hence on time. Nevertheless, $\beta_{l,m}$ could fluctuate
around an average value $\langle \beta_{l,m} \rangle$ in such a rapid fashion 
as to be unresolved on macroscopic scales, making the macroscopic 
behaviour of the GNS model identical to one with given constant viscosity. 
This is our interpretation of the EC of \cite{Gal97} which for 
$l=1, m=0$ states that:

\vskip 5pt
\noindent
{\bf Equivalence Conjecture (EC).} {\it The stationary probability 
distributions of the NS equations and of the GNS equations are 
{\em equivalent} in the limit of large Reynolds number, provided the 
enstrophy $Q$ and the viscosity $\nu$ are so related that $\sigma_l^{NS}$ 
and the average of $\sigma_{l,m}^{GNS}$ are equal.}

\vskip 5pt
\noindent
In Ref.\cite{Gal97} the large Reynolds number plays the role
of the thermodynamic limit, in which limit $\beta_{l,m} \zs_l^{NS}/\nu_l$ in
Eq.(\ref{sigmarev}) is expected to dominate over the other terms, making the
equality $\zs_l^{NS} = \langle \zs_{l,m}^{GNS} \rangle$ imply 
$\langle \beta_{l,m} \rangle= \nu_l$, for $l=1, m=0$. 
We further conjecture that the same should hold for any $l$ and $m$.

How can these reasonings be justified?  In the first place the
solutions to the NS equations  are expected to take a form, in the
$\nu_l\to 0$ limit, which is independent of the viscosity and of the
boundary conditions, and therefore universal.  This is because the
flow is expected to develop on characteristic length scales which are
much smaller than the system size but much larger than the small scale
at which the dissipation is effective.  This limit is then natural 
in hydrodynamics, as it is desirable for the EC to
have a universal rather than particular character. Moreover, in
standard computational practice, the limit $\nu_l \to 0$ is connected
to the $N \to \infty$ limit: a numerical simulation of turbulence is
considered reliable when the dissipative scale $1/k_d$ is well
resolved ($N>L k_d$, $L$ being a length scale of the system). The
value $k_d$ is estimated by applying Kolmogorov's theory, which
assumes a power law scaling in a range of $k$ called inertial; this
scale becomes smaller as $\nu_l\to 0$. However, the GNS case is more
garbled, as it is not known how the scaling is affected by the
truncation and by the reversible dissipation.  Arguments based on
power counts shed little light on this, at least for the easier cases
in which $R_{m,l}=0$: for increasing $N$, $\sigma_\nu^{NS}$ behaves as
\be
  \sigma_l^{NS} = \nu_l \sum_{k_x=-N}^N\sum_{k_y=-N}^N \kk^{2l}\sim
  2 \nu_l \pi \int_1^N k^{2l+1} dk \sim \nu_l \frac{N^{2l+2}-1}{2l+2}\, .
\ee
In order to derive scaling relations in $N$ for the quantities involved 
in the definition of $\sigma_{l,m}^{GNS}$, we can assume that the shell energy
$E(k)$, defined as
\be
  E(k)=\sum_{[|\kk|]=k}\frac{|\omega_{\kk}|^2}{\kk^2}\, ,
\ee
where $[\cdot]$ denotes the nearest integer, scales in $k$ as $k^{\lambda}$. 
Such a power law provides at least a
standard term of comparison. The reference theory for $2D$ turbulent flows,
due to Batchelor, Kraichnan and Leith \cite{Krai67}, predicts
$\lambda=-3$. This slope would appear only in the inertial range, and
a steeper decay of $E(k)$ is expected in the dissipative range
$k>k_d$. It can then be inferred that, for $N$ large, but smaller 
than $L k_d$, $Q_n$ scales in $N$ as
\be
  Q_n=\sum_{k_x=-N}^N\sum_{k_y=-N}^N \kk^{2n}|\omega_{\kk}|^2 
  \sim \sum_{k} k^{2n+2} E(k) \sim 2 \pi \int_1^N k^{2n+\lambda+2} dk 
  \sim \frac{N^{2n+\lambda+3}-1}{2n+\lambda+3}\,.
\ee
This implies that $Q_n$ converges for large $N$ if $\lambda<2n+3$, while
it would diverge in the opposite case. This divergence might indeed not
be observed, as the limit 
$N\to\infty$ has to be taken before $\nu_l\to 0$, i.e.\ a dissipative range
is always included. As for $F_m$ in Eq.(\ref{revdissk}), we argue that for a
given forcing concentrated on a few
(or just one) $\kk$ vectors, the term $F_m$ can depend on $N$ only if 
the magnitudes of $|\omega_\kk|$ or the phase between $\omega_\kk$ and $f_\kk$
change when adding more ultraviolet modes. If we assume that the (large scale)
vorticty field is not significantly altered by the (much smaller) additional
large $N$ modes, then also $F_m$ should not change significantly while
increasing $N$. Eq.(\ref{revdissk}) therefore says that
\be
  \beta_{l,m}\sim\frac{1}{Q_{l+m}} \sim \frac{1}{|N^{2m+2l+\lambda+3}-1|}\,,
  \qquad m=-1,0 \,.
\ee
Our results reported in Section 4 show that
$\beta_{l,m}\simeq\nu_l$, for all $N$, so we should have
$\lambda<-2m-2l-3$ (e.g.\ $\lambda<-3$ for constrained energy and $\lambda<-5$
for constrained enstrophy and normal viscosity), provided that
$E(k)\sim k^\lambda$ and within the inertial range. With this constraint
for $\lambda$, we have
\be
  \frac{Q_{2l+m}}{Q_{l+m}} =  \left\{ 
\begin{array}{ll}
{\mathcal O}(N^{2l}) ~~\mathrm{if }~ 0>\lambda+2m+2l+3>-2l \\ 
{\mathcal O}(N^{0}) ~~\mathrm{if }~   \lambda+2m+2l+3<-2l\, 
\end{array}
\right.
\ee
The power count in eq.\ (\ref{sigmarev}) then shows that for $l\ge0$ 
we have
\be
  \sigma_{l,m}^{GNS} = 
  \frac{\beta_{l,m}}{\nu_l} {\mathcal O}(N^{2l+2}) \,,
   \label{ordNsig}
\ee
so that for large $N$ (depending on $\nu_l$) $\sigma_{l,m}^{GNS}$ is 
proportional to
$\beta_{l,m}$, in the leading order. Indeed this is correct for any $N$, if
$l=0$, and numerical inspection shows that, in all our simulations (made with
$l=1$ and $m=-1,0,1,2$), the ratio
$\nu_l\langle\sigma_{l,m}^{GNS}\rangle/(\sigma_l^{NS}\langle\beta_{l,m}\rangle)$ differs
from one by no more than a few parts per thousand, {\it already at truncations 
with $N$ as low as 3}. We can therefore fix values of $\nu_l$ and number of
modes $N$, and check the equality $\langle \beta_{l,m} \rangle = \nu_l$
needed for the macroscopic equivalence of the NS and GNS systems, from
which the equality $\zs_l^{NS} = \langle \zs_{l,m}^{GNS} \rangle$ follows.

Whether we take the thermodynamic limit or not, there is a formal similarity
between the cut-off GNS equations and Eqs.(\ref{eqsmot}), which may then enjoy
similar properties. Of these, the validity of the GCFT of \cite{GalCoh95} is
perhaps the most striking one, and in \cite{Gal97} Gallavotti has argued that a
similar relation (introduced in \cite{BoGaGa96})
should hold for the fluctuations of $\zs_{l,m}^{GNS}$,
defined by Eq.(\ref{sigmarev}) with $l=1$ and $m=0$. Because of 
Eq.(\ref{ordNsig}), a similar fluctuation law should also hold for the 
fluctuations of $\zb_{l,m}$ (possibly for any $l,m$), which has a more direct 
physical interpretation than $\zs_{l,m}^{GNS}$, since $\zb_{l,m}$ represents  
the viscosity in the GNS equations. 

To test these ideas in our simulations of a $2D$ fluid, we need the following
definitions. Let $t \mapsto V_t \zw$ be the time evolution for an initial
vorticity field $\zw$, take an overall simulation time $T$ which must be 
adequately longer than the characteristic time of the fluctuations of 
$\beta_{l,m}$, and assume that relaxation to a statistically stationary
state takes place in a time $t_0 \ll T$. Then, introduce the time average
\be
  \langle \beta_{l,m} \rangle = \frac{1}{T} \int_0^T
  \beta_{l,m}(V_t \zw) d t \, .
\ee
We omit to indicate the dependence of 
$\langle \beta_{l,m} \rangle$ on $T$ and on $\zw$, assuming that $T$ is 
large enough, and the dynamics ergodic enough, that larger simulation 
times and different randomly chosen initial fields would produce 
negligible differences in the end result. Let us subdivide the time 
interval $[t_0,T]$, in a number of subintervals of length $\zt$, and
consider the quantities
\be
  \overline{\beta}^{\;\tau}_{l,m}(i) =
    \frac{1}{\langle\beta_{l,m}\rangle \tau} \int_{t_0+(i-1)\zt}^{t_0+i\tau}
    \beta_{l,m}(V_t \zw)~d t ~; \quad i = 1, ..., \frac{T-t_0}{\tau}~.
\ee
These values, arranged in a histogram, allow us to construct the probability 
distribution $\pi_{l,m}^\zt$ of the fluctuations of $\beta_{l,m}$ integrated
over a time $\zt$. Then, a test of the validity of the fluctuation relation
Eq.(4.1) of \cite{Gal97} amounts in our setting to verify that the values 
\be
  C_\beta^{l,m}(\zt,p) = \frac{1}{\tau \langle \beta_{l,m} \rangle p} 
  \log \frac{\pi_{l,m}^\tau (p)}{\pi_{l,m}^\tau (-p)} \, ,
  \label{FluTh} 
\ee
obtained for different values of $\zt$ and $p$ (in the support of 
$\pi_{l,m}^\tau$), do converge for growing $\tau$ to a definite value 
$c_\beta^{l,m}$ independent of $p$. 
The fluctuation relation of \cite{Gal97}
is then a statement on the form of the measured p.d.f.\ $\pi_{l,m}^\tau$ 
for large $\tau$. We found that $C_\beta^{l,m}(\zt,p)$ can 
be fitted by a $p$-independent function of $\zt$, $C_\beta^{l,m}(\zt)$ say. 
Therefore, the odd part of $\log(\pi_{l,m}^\tau(p))$ appears to be linear 
in $p$ at all $\zt$. Then, to assess the validity of
the fluctuation relation, we verify whether the slope 
$C^{l,m}_\beta(\tau)$ of the linear 
function of $p$
\be
  \frac{1}{\tau\langle\beta_{l,m}\rangle} \left[
  \log \left( \pi_{l,m}^\tau(p) \right) - 
  \log \left( \pi_{l,m}^\tau(-p) \right)
  \right] = C^{l,m}_\beta(\tau) p 
\label{finite}
\ee
does converge to a given $c_\zb^{l,m}$, as $\zt$ is increased. This 
linear form is shared by many kinds of distributions, including distributions 
with exponential tails and Gaussian distributions, giving the erroneous
impression that a trivial connection exists between the fluctuation relations 
of \cite{GalCoh95,Gal97} and the central limit theorem (see point 4 of 
Section 5).

\section{Description of the numerical method}
We performed a number of numerical experiments, integrating in time
several vorticity fields and using a standard protocol for the NS
equations \cite{CaHuQuZa88}. We used a two-dimensional $2/3$ dealiased
pseudospectral code on the periodical square $[0,2\pi] \times
[0,2\pi]$. The resolution of such simulations is traditionally
expressed by the number of modes $M\times M$ over which the
convolution term is computed; the number $M$ is often a power of 2 for
FFT convenience. With the dealiasing procedure, the vorticity field is
reconstructed in physical space using only the modes with $-M/3\le k_x,
 k_y \le M/3$. The formulas given above have therefore to be read
with $N=M/3$: for example, $M=32$ implies a system of 221 complex
coupled modes.  Time advancement in our algorithm is accomplished by a
fourth order Runge--Kutta integrator. This NS code is easily modified
to integrate the GNS equations; even the convolution term $R_{m,l}$ in
(\ref{Rmlkpq}) can be computed via  a multiplication in physical
space, as is done according to the pseudospectral method for the
nonlinear term $n_k$ of eq. (\ref{nltk}).  However, in the GNS case
the analytical integration of the linear viscous term, often done to
allow longer integration steps for the NS equations, is not
possible. Moreover, the mere substitution of the constant $\nu_l$ by
$\beta_{l,m}$, to be computed at each substep of the
Runge--Kutta algorithm, does not guarantee the exact conservation of
the constrained global quantity $Q_m$. This is not a real problem,
because the relative error  $\epsilon =
\left|Q_m(t_0)-Q_m(t_0+T)\right|/Q_m(t_0)$ made in the conservation of
$Q_m$ can be controlled by an adequate reduction of  the integration
time steps. In Table \ref{runtab}, the values of $\epsilon$  for each
run are explicitly given, confirming that our accuracy in the
conservation of $Q_m$ is satisfactory.

\begin{figure}
\epsfysize=9cm 
\epsfxsize=14cm
\centerline{\epsfbox{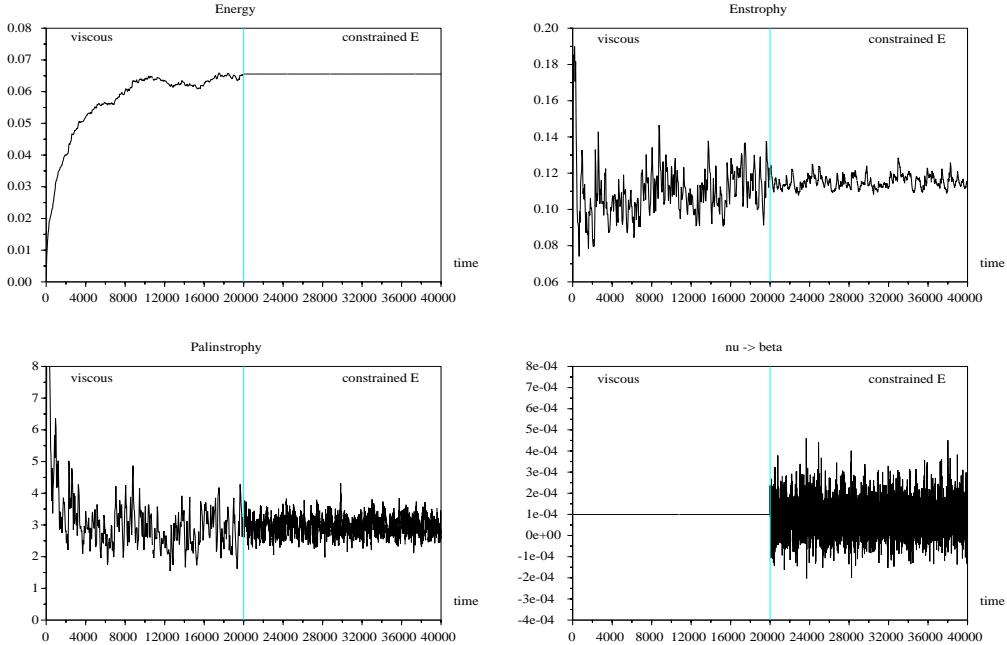}}
\caption{Run \#9: behaviour of $E$, $Q$, $P$ in the NS ``buildup'' part
    (first halves of the plots), followed by the behaviour of the same 
    quantities and of $\beta_{1,-1}$ once $E$ is constrained according to
    the GNS dynamics (second halves of the plots).
    Data points are subsampled for graphical pourposes. 
    $\langle \beta_{1,-1} \rangle$ is close to $\nu_1$.
   \label{acc32ME}}
\end{figure}

\begin{figure}
\epsfysize=9cm \epsfxsize=14cm
\centerline{\epsfbox{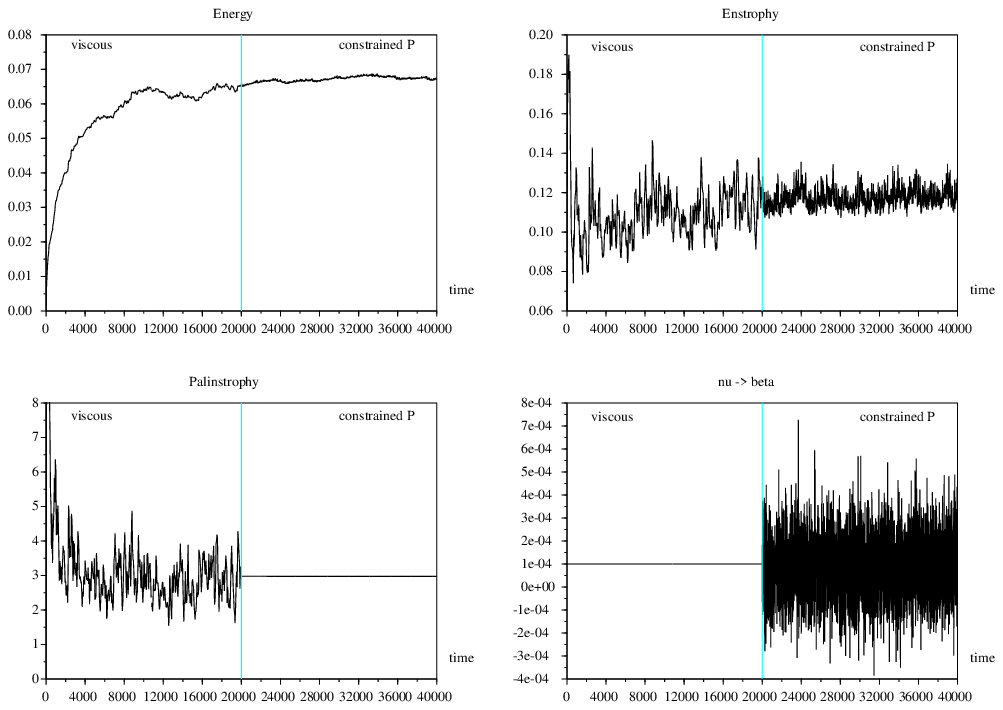}}
\caption{Run \#11: behaviour of $E$, $Q$, $P$ and $\beta_{1,1}$ in the NS 
   ``buildup'' part and once $P$ is constrained.
    The format of presentation is the same of Fig.\ref{acc32ME}.
    Also here $\langle \beta_{1,1} \rangle$ is close to $\nu_1$.
   \label{acc32MP}}
\end{figure}

To study the properties of statistically stationary, nonequilibrium
states both for the NS and for the GNS systems, we need a criterion to
judge if a stationary state has been reached in a simulation. We found
that, for the NS system, the approach to a steady state can be
assessed by looking at the time series of the quantity $Q_m$ which is
going to be constrained in the GNS system. Fluctuations of such
quantities in the NS system have amplitudes and correlation times
which vary from case to case, but they take relatively short times to
settle around  an average value, if $\nu_l$ is not too small and the
forcing is not too  large.\footnote{This and the fact that sufficient
numerical accuracy requires very small integration time steps if
$\nu_l$ is small and $f$ large, determines the range of
parameters  which we can investigate.} As an empirical criterion, we
then assume that the  NS system has reached a stationary state when
$Q_m$ fluctuates around a  definite value. Figures \ref{acc32ME} and
\ref{acc32MP} show how this is typically achieved.

To perform our NS simulations, we have first to choose the values of
the viscosity coefficient $\nu_l$, of the force $f_\kk$, and of the
number of active modes,  i.e.\ of $N$. The correspondig GNS system is
obtained keeping the same $f_\kk$ and $N$, discarding $\nu_l$, and
choosing the global quantity to be fixed which, in our simulations,
was $E$ or $Q$ or $P$ or $H$. Once this is done, we  let the NS system
evolve towards a steady state using a ``buildup'' approach:  starting
from a small random initialization of the $\W$ modes, we numerically
integrate the field for a sufficiently long time, $t_0$
say. Simulations are  considered sufficiently long in this context,
when they are long with respect  to the correlation times of the
fluctuating series and to the time  required to stabilize the average
of the considered $Q_m$.  A snapshot of the vorticity field at an
arbitrary time $t>t_0$, is then used as initial datum for the GNS runs.

In order to perform longer runs, we minimized the number of modes  of
the vorticity, i.e.\ the resolution. Some numerical experiments led us
to conclude that, given a force $f_{\kk}$, an adequate resolution
should include the spectral modes up to at least $2\kk$. Thus, for
instance,  we performed many of our runs with forcings on $\kk=(\pm
3,\pm 4)$ and  $\kk=(\pm 4,\pm 3)$  at a resolution of $M=32$, which,
taking into account the dealiasing, implies a truncation of the
Fourier expansions at $N=10$. Such resolution allows us to carry on
simulations for up to some million timesteps, i.e. to total simulation
times $T$ of the order of $10^5$ time units, whereas fluctuations show
characteristic times of the order of ten units. Runs at higher
resolution converge to a steady state with approximately the same
values of the chosen $Q_m$, in roughly the same time and, in general,
with similar fluctuation amplitudes. 
The values of the Reynolds numbers (computed as $\frac{E}{\nu_1 \sqrt{Q}}$)
are not particularly relevant in our context; however, they are observed
to be of the order of $10^2$ up to $10^4$, in the various runs.
The complete list of runs
performed, inclusive of simulation  parameters and results, is
presented in Table \ref{runtab}.

\begin{table}
\centerline{
\begin{tabular}{r|r|r|r|r|r|r|r|r|r}
 \hline 
run &     & integ.     & total $T$      & $Q_m$   & constraint & force      
 & $\nu_1$ & & asymptotic   \\
\#  & $M$ & $\Delta t$ & $\times 10^3$  & fixed  & error $\epsilon$ & $f$
 & $\times 10^4$  & $\langle\beta_{1,m}\rangle/\nu_1$ & $c_\sigma^{1,m}$ \\
 \hline
1 & 32 & 0.1 & 100 & E & $3\cdot 10^{-10}$ & $f_{55}$
 & $1$ & $\bf 0.9543$ \\ \hline 
2 & 32 & 0.1 & 100 & E & $3\cdot 10^{-7}$ & $10 f_{55}$
 & $1$ & $\bf 1.0472$ \\
3 & 32 & 0.1 & 490.178 & Q & $1\cdot 10^{-4}$ & $10 f_{55}$
 & $1$ & $\bf 1.0654$  \\
4 & 32 & 0.1 & 100 & P & $8\cdot 10^{-4}$ & $10 f_{55}$
 & $1$ &  $\bf 0.9399$ & $0.29 \pm 0.03$  \\
5 & 32 & 0.1 & 200 & H & $2\cdot 10^{-4}$ & $10 f_{55}$
 & $1$ & $\bf 0.9024$ & $0.17 \pm 0.09$ \\
6 & 256 & 0.1 & 5 & E & $4\cdot 10^{-9}$ & $10 f_{55}$
 & $1$ & $\bf 1.1276$   \\
7 & 256 & 0.1 & 5 & Q & $6\cdot 10^{-6}$ & $10 f_{55}$
 &  $1$ & $1.6818$\\
8 & 256 & 0.1 & 5 & P & $2\cdot 10^{-6}$ & $10 f_{55}$
 & $1$ & $1.6024$ \\ \hline 
9 & 32 & 0.025 & 150 & E & $5\cdot 10^{-5}$ & $100 f_{55}$
 & $1$ & $\bf 0.9497$ & $0.30 \pm 0.01$ \\
10 & 32 & 0.025 & 150 & Q & $4\cdot 10^{-3}$  & $100 f_{55}$
 & $1$ & $\bf 0.8891$  & $0.35 \pm 0.03$ \\
11 & 32 & 0.025 & 125 & P & $1\cdot 10^{-2}$ & $100 f_{55}$
 & $1$ & $\bf 0.9344$ & $0.27 \pm 0.04$ \\ 
12 & 64 & 0.025 & 25 & E & $9\cdot 10^{-6}$ & $100 f_{55}$
 & $1$ & $\bf 1.0189$ & $0.011 \pm 0.001$ \\
13 & 64 & 0.025 & 25 & Q & $5\cdot 10^{-3}$ & $100 f_{55}$
 & $1$ & $\bf 1.0359$ & $0.0188 \pm 0.0006$ \\
14 & 64 & 0.025 & 25 & P & $3\cdot 10^{-2}$ & $100 f_{55}$
 & $1$ & $\bf 0.9837$ & $0.075 \pm 0.01$ \\ \hline 
15 & 32 & 0.002 & 2  & E & $5\cdot 10^{-8}$ & $1000 f_{55}$
 & $1$ & $\bf 0.9439$ & $0.37 \pm 0.03$ \\
16 & 32 & 0.002 & 2  & P & $1.5\cdot 10^{-5}$ & $1000 f_{55}$
 & $1$ & $\bf 1.02558$ & $0.19 \pm 0.02$ \\ \hline
17 & 32 & 0.025 & 150  & E & $8\cdot 10^{-7}$ & $100 f_{55}$
 & $3$ & $\bf 0.9406$ & $0.23 \pm 0.05$ \\
18 & 32 & 0.025 & 150  & Q & $2\cdot 10^{-4}$ & $100 f_{55}$
 & $3$ & $0.7553$ & $0.36 \pm 0.08$\\
19 & 32 & 0.025 & 150  & P & $9\cdot 10^{-6}$ & $100 f_{55}$
 & $3$ & $1.6349$ & $\sim 0.2$ \\ \hline 
20 & 32 & 0.025 & 150  & E & $3\cdot 10^{-4}$ & $100 f_{55}$
 & $0.5$ & $\bf 1.045$ & $0.31 \pm 0.02$ \\
21 & 32 & 0.025 & 150  & Q & $2\cdot 10^{-2}$ & $100 f_{55}$
 & $0.5$ & $\bf 1.0648$ & $0.409 \pm 0.003$ \\ \hline
22 & 12 & 0.1 & 2.614  & E & $9\cdot 10^{-8}$ & $f_{22}$
 & $1$ & $\bf 1.0052$ \\
23 & 32 & 0.1 & 10 & E & $4\cdot 10^{-6}$  &$10 f_{5a}$
 & $1$ & $\bf 1.0260$ \\
24 & 32 & 0.1 & 10 & E & $8\cdot 10^{-8}$& $f_{2c}$
 & $1$ & $0.7266$ \\
25 & 32 & 0.1 & 10 & E & $4\cdot 10^{-5}$ & $10 f_{5a} + f_{2c}$
 & $1$ & $\bf 0.9694$ 
\end{tabular}}
\caption{Runs performed. Here $\Delta t$ represents the integration step, 
$T$ the total integration time and $\epsilon$ the relative
numerical error on the constrained quantity $Q_m$.
The following notation is also used:
$f_{55}=f_{5a}+f_{5b}$, with $f_{5a}=0.00015$ on ${\bf k}=(4,-3)$ and
$f_{5b}=0.0001 i$ on ${\bf k}=(3,4)$; $f_{22}=f_{2a}+f_{2b}$, with
$f_{2a}=0.0006$ on ${\bf k}=(2,0)$ and $f_{2b}=0.0004 i$ on ${\bf k}=(0,-2)$;
$f_{2c}=-0.00009+0.00008i$ on ${\bf k}=(0,-2)$. Values of
$\langle\beta_{1,m}\rangle/\nu_1$ which agree with the (extended) EC 
within $15\%$ error are marked in boldface, where it is intended
that an error may affect the last digit. The values
$c_\sigma^{1,m}$ and the associated errors were computed 
by fitting the final part of 
curves like those in Fig.\ref{sloconv} with horizontal lines,
when sufficient statistics was available. 
\label{runtab}}
\end{table}

\section{Tests: equivalence conjecture and fluctuation formula}
Our GNS systems, with our choice of  parameters, take a reasonable
computation time to reach a stationary  state, i.e.\ a state in which
the fluctuations of $\beta_{l,m}$ occur  about its time average
$\langle \beta_{l,m} \rangle$.  This quantity is found to coincide
(apart from small errors, and from the cases with insufficient
statistics) with the value  of the viscosity $\nu_l$ of the NS
equations, if the quantities $E, Q, P$ and  $H$ are fixed to the value
measured in the final NS snapshot. This fact is quite  remarkable if
we consider that the only information we pass from the NS to the  GNS
system is the NS field at an arbitrary time $t$ after the stationary
state  has been reached, and that two consecutive snapshots may yield
rather different  values of the chosen $Q_m$. In doing so, we assumed
that any phase space point  of the steady state trajectory carries
enough information for the GNS run, and  our results validate this
approach.

The fact that good agreement is obtained when imposing the
conservation of a quantity like $E$ or $Q$ is not entirely surprising,
and  might be connected with the notion of ``rugged invariants'',
i.e.\ of global quantities which are approximately conserved in
spontaneous turbulent decay. While the constancy of such quantities
has been imposed for structural reasons here, that was earlier the key
assumption of a number of theories of turbulence. Without attempting
to make a review, we remark that the stationarity of $E$, together
with that or not of $Q$, were the basis of various models of either
stationarily forced or freely decaying two-dimensional
turbulence. Among these we quote: a) the equilibrium statisical
mechanics of Fourier modes, with fixed $E$ and $Q$, due to Kraichnan
\cite{KraMo80}; b) the minimum enstrophy theory, in which the decay is
ruled by a minimization process of $Q$ at fixed $E$ \cite{Leith84}; c)
the maximal entropy theory, in which the decay involves again a
lowering of $Q$ at constant $E$, but the decrease in $Q$ is explained
probabilistically as loss of information due to mixing
\cite{MiWeCr92}.  In our cases, however, the hierarchy of conserved
quantities implied in these theories seems to be absent: the
equivalence between the stationary state of the GNS and
of the NS equations is found for all constrained $Q_m$. On the other
hand, the amplitude of the fluctuations of $\beta_{l,m}$ increases
with $m$, as can be seen comparing Figs.\ref{acc32ME}  and
\ref{acc32MP}.

\begin{figure}
\centerline{\epsfxsize=12cm \epsfysize=9cm \epsfbox{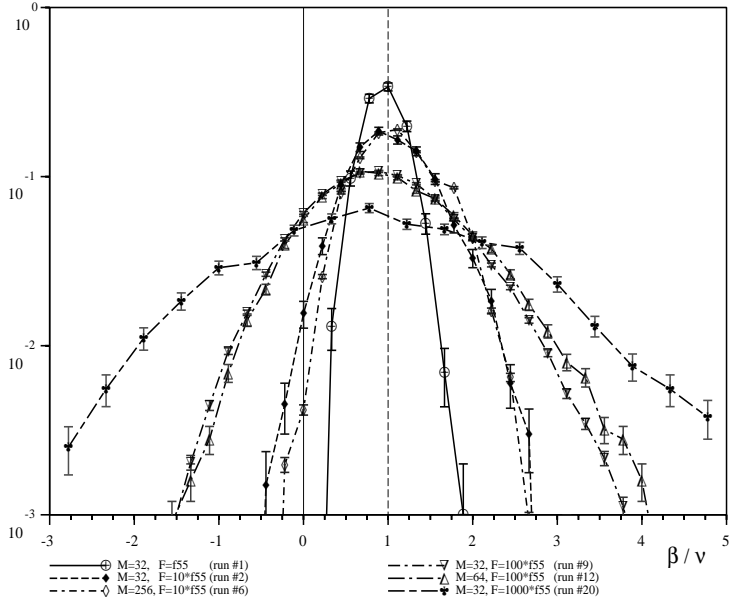}}
\caption{Histograms of $\beta_{1,-1}$ with error bars, for runs with
constrained $E$, original $\nu_1=10^{-4}$ and various forcings.  Note
the coalescence of histograms referring to runs at  different
resolution but equal forcing.
\label{histFE}}
\end{figure}

Another interesting fact is that in the NS equations we can fix the
value of $f$ and of $\nu_l$ independently of each
other, while the amplitude of the fluctuations of $\beta_{l,m}$ grows
as the ratio $f/\nu_l$ increases (see Fig.\ref{histFE}). This is
different from the case of particle systems, where the dissipative
term is proportional to the forcing term. Also, differently from the
case of particle systems, the quantity $\dot{\omega}_k$ in the cut-off
GNS equations has $k^{2l} \beta_{l,m}\omega_k$ instead of just
$\beta_{l,m} \omega_k$. It is striking, then, that the equality
between $\langle \beta_{l,m} \rangle$ and $\nu_l$ is verified
irrespective of the value of $f$,  as seen varying $f$ over several
orders of magnitude.

\begin{figure}
\centerline{\epsfxsize=10cm \epsfysize=7cm \epsfbox{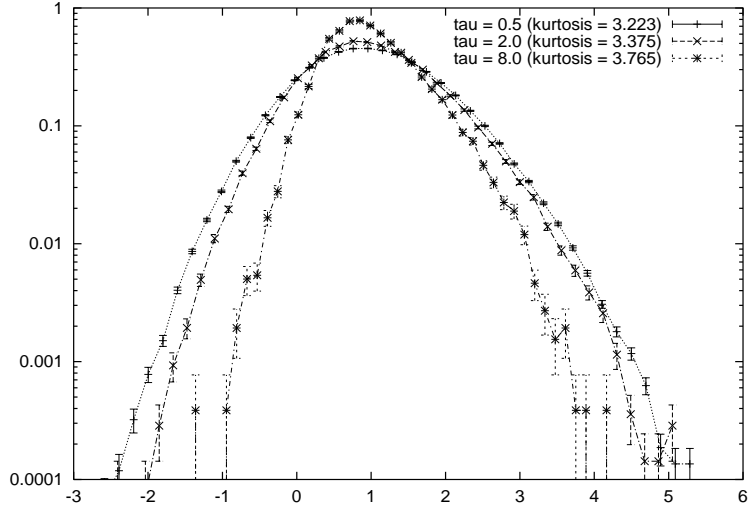}}
\caption{Normalized histograms of $\overline{\beta}^{\;\tau}_{1,1}$
for various values of $\tau$,  for the run \#9. The values of the
kurtosis evidence the non-Gaussian nature of these distributions,
which becomes  more and more pronounced as $\tau$ grows.
\label{betaden}}
\end{figure}

To check the validity of the fluctuation relation
Eq.(\ref{finite}), we considered a number of long runs,
which we cut in segments of given length $\tau$, for various values of
$\tau$, for different realizations of the forcing term, and for
different values of the viscosity $\nu_1$. Moreover, we
considered four  different constraints in the cut-off GNS equations:
$E$, $Q$, $P$ or $H=$constant. Sample histograms of the distribution
of the averages of $\overline{\beta}^{\;\tau}_{1,-1}$ over the times
$\tau$ are shown in Fig.\ref{betaden} for constant $E$ and a few
different values of $\tau$. Clearly, the distributions become more and
more peaked around the mean $\langle \beta_{1,-1} \rangle$ as $\tau$
is increased (large fluctuations of $\beta_{1,-1}$ are wiped out by
the average).  Such distributions deviate however from gaussian, and
the non-gaussian character is not mildened by the averaging; the
actual computation shows that the kurtosis
$\langle(\overline{\beta}^{\;\tau}_{l,m}-1)^4\rangle /
\langle(\overline{\beta}^{\;\tau}_{l,m}-1)^2\rangle^2$ is
significantly  different from 3 (the value for the gaussian
distribution), and that this  difference seems to increase both with
$\tau$ and $f$ (see Fig.\ref{betaden}).  This is not exactly the case
for particle systems (at moderate forcings),  whose probability
distributions can be interpolated quite well by Gaussians, although
they should not be Gaussian distributions \cite{BoGaGa96}. 
Similarly to previous  works
\cite{EvCoMo93,BoChLe98} for particle systems, we do not separate in
time  the adjacent segments to decorrelate them. The reason is that,
in all the cases we considered, the only effect of the decorrelation
was the worsening of data statistics. Our test consisted in computing
the left hand  side of Eq.(\ref{finite}) for many values of $p$, in
verifying the consistency  of the results with a linear law of slope
$C_\beta^{l,m}(\tau)$, and in extrapolating  
$C_\beta^{l,m}(\tau)$ to large $\tau$. Given the observed  proportionality
between $\beta_{l,m}$ and $\sigma_{l,m}^{GNS}$, Eq.(\ref{finite})
holds for the distributions of $\sigma_{l,m}^{GNS}$,  
with the slope $C_\beta^{l,m}(\zt)$ replaced  by
$C_\sigma^{l,m}(\zt) = C_\beta^{l,m}(\zt) \nu_l / \sigma_l^{NS}$.

\begin{figure}
\centerline{\epsfxsize=10cm \epsfysize=7cm \epsfbox{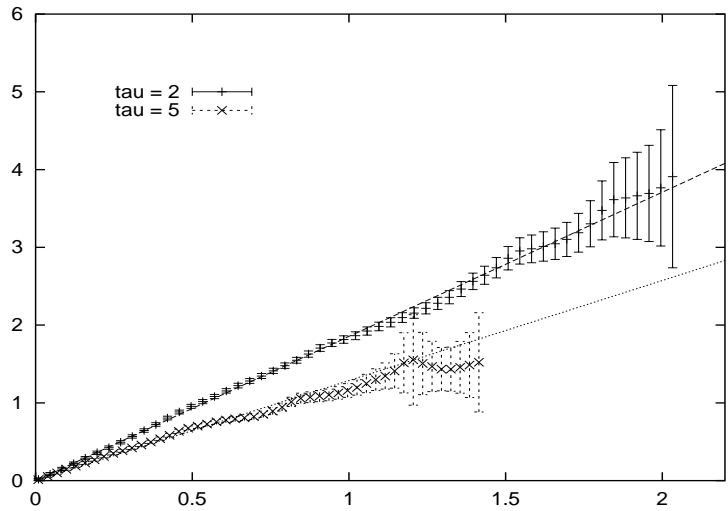}}
\caption{Linear fit of  $\frac{\log\pi_{l,m}^\tau(p)-\log\pi_{l,m}^\tau(-p)} {\tau
\langle\beta_{1,-1}\rangle}$ to $C^{1,-1}_\beta(\tau)\;p$  for two
different averaging times $\tau$, from run \#9.
\label{linfit}}
\end{figure}

\begin{figure}
\centerline{\epsfxsize=14cm \epsfbox{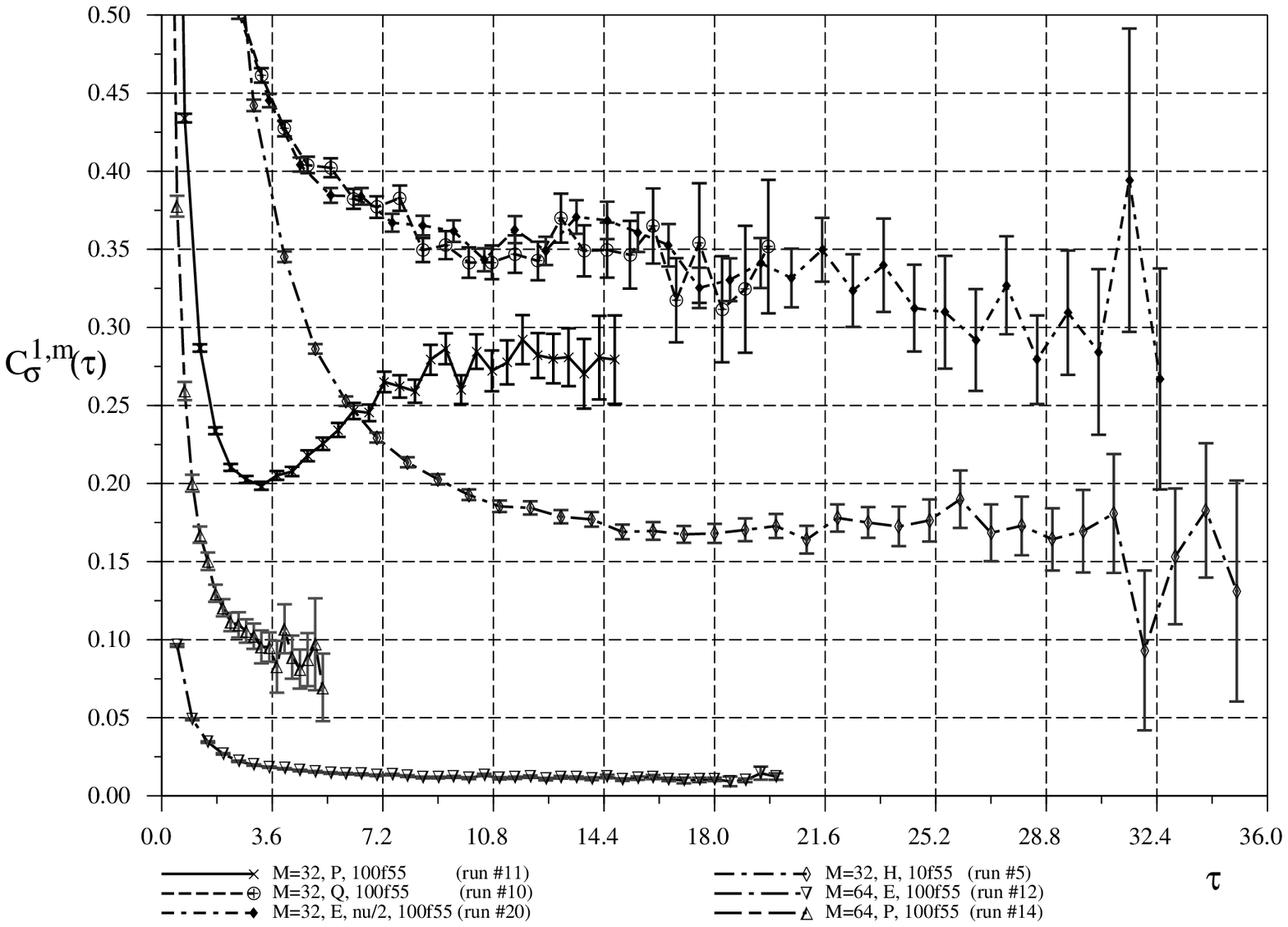}}
\caption{Convergence of the slope $C_\sigma^{1,m}(\tau)$ of the
regression line  for various runs. The legend stresses the resolution
M, the quantity constrained, the force used and the parent viscosity,
also reported in Table \ref{runtab}. The data presented here refer
respectively to the runs \#11, \#10, \#20, \#5, \#12 and \#14 of that
table. Note the overlap of the curves of runs \#10 and \#20. 
\label{sloconv}}
\end{figure}

Our results show that the  distributions of the values
$\overline{\beta}^{\;\tau}_{l,m}$ are in all cases  consistent with
the linear law Eq.(\ref{finite}). The fit to a straight  line,
performed according to standard least-squares minimization, is always
remarkably good, as in the cases shown in Fig.\ref{linfit}. The error
bars are derived from the errors on the histograms, which, according
to standard statistical analysis, are estimated as the square root of
the bin count. The value of the linear regression coefficient vs.\
$\tau$ is plotted in Fig.\ref{sloconv} for several cases. The slope of
these lines decreases with $\tau$, and converges to a finite value at
large $\tau$ when $E$, $Q$, $P$ or $H$ are fixed. Therefore, the
fluctuation relation is valid  in these cases, although presently we
cannot compare our results with  Gallavotti's theoretical prediction
\cite{Gal97} which links $c_\sigma^{l,m} =
\lim_{\tau\rightarrow\infty} C_\beta^{l,m}(\tau) \nu_l / \sigma_l^{NS}$ to the
Lyapunov  spectrum of the system at hand. However, consistently
with these predictions, we obtain $0 < c_\zs^{l,m} < 1$. 
We conclude this section
noting that the  dependence of $C_\beta^{l,m}(\tau)$ on $\tau$ (cf.\
Fig.\ref{sloconv}) is different from that observed in the work
\cite{CL}, while it is similar  to that observed in the particle systems of
Ref.\cite{BoChLe98}.

\section{Conclusions}
{\bf 1.} We investigated the equivalence between the NS and GNS
equations. Such equivalence is verified for a variety of cases, and
actually appears more  robust than expected in \cite{Gal97}. Indeed,
$\langle \beta_{1,m} \rangle$  approaches $\nu_1$ at least when $E$ or
$Q$ or $P$ or $H$ are fixed in the  cut-off GNS equations (cf.\ Table
\ref{runtab}). Moreover, neither the thermodynamic limit nor a large
number of modes is needed to obtain this result.  This leads us to
conclude that the EC can be extended to the case of  constrained
$Q_m$, for any $m \ge -1$. The equivalence  of the NS and GNS
equations can be equally expressed in terms of $\langle  \beta_{l,m}
\rangle$ and $\nu_l$, or of $\langle\zs_{l,m}^{GNS}\rangle$ and $\zs_l^{NS}$,
due to the proportionality between $\zs_{l,m}^{GNS}$ and $\zb_{l,m}$
expressed by  Eq.(\ref{ordNsig}).  We preferred to use
$\langle\beta_{l,m}\rangle$, which is directly connected  with a
quantity of ordinary hydrodynamic interest: the viscosity of the fluid
$\nu_l$. Moreover, the values of $c_\beta^{l,m}$ appear
to cluster  around a single value, independent of the resolution of
the run and of the  parent viscosity, as can be inferred from Table \ref{runtab} using $c_\sigma^{l,m} = c_\beta \nu_l / \sigma_\nu^{NS}$. 
This is not the case for $c_\sigma^{l,m}$.

{\bf 2.} The fluctuation relation
Eq.(\ref{finite}) is seen  to hold when $E, Q, P, H$
are constrained. In these cases, the left hand side of
Eq.(\ref{finite}) is perfectly fitted by straight lines at all values
of $\zt$, (see e.g.\ Fig.\ref{linfit}) with slopes converging to
definite values $c_\sigma^{l,m}$ in the limit of large  $\zt$ (cf.\ Fig.\ref{sloconv} and Table \ref{runtab}). This at one time validates the 
CH in the framework of hydrodynamics, and yields  a new result related to 
the NS equations. Moreover, the values $c_\sigma^{l,m}$
are consistent with Gallavotti's predictions (cf.\ Eq.(4.1) of \cite{Gal97}), which imply $0 < c_\sigma^{l,m} \le 1$.  

{\bf 3.} Our results concern statistically stationary states of the
dynamics  which, from a fluid--dynamicist point of view, are achieved
at very late times. The resulting flows are substantially dominated,
in absence of a proper large-scale dissipation  mechanism, by
vorticity structures of the size of the computational domain.  These
large scale structures are seen to be unsteady, fact which is by
itself responsible for the fluctuations of the system.  In such
conditions the mechanism of transfer of energy toward small  $k$ has
already pushed almost all the avaliable energy to the largest
accessible scales; the formation of a self similar direct enstrophy
cascade or of an inverse energy cascade are prevented. If one is
concerned with universal inertial range properties of the turbulence,
such kind of states would be inappropriate because they are dependent
on the finite size boundary conditions, hence not universal.  Our
point of view is indeed different, and we do not consider the poor
resolution or the lack of a universal turbulent cascade as
shortcomings. We are interested in the statistical behavior of a given
dynamical system, and the computational limitations on $N$ do not
hinder our investigation. In this spirit, we treated the viscosity
coefficient as a mere parameter of the simulation; we did not match
resolution and viscosity in order to guarantee the resolution of a
dissipative range. Again, the point is that we investigate special
properties of the vorticity modes which, because of their dynamical
nature, should not depend on the level of resolution.

{\bf 4.} Some confusion is sometime made on the nature of the GCFT  of
Ref.\cite{GalCoh95}, because the p.d.f.\ obtained for the fluctuations
of the entropy production rate in particle systems are
indistinguishable to the eye from Gaussian distributions.  This led
some to believe that the statement of a linear law for the  odd part
of the p.d.f.\ of the fluctuations in particle  and hydrodynamic
systems is a quite generic result and just a manifestation of  the
validity of the central limit theorem. This belief, however, is not
well founded. In the first place, the relation investigated here and
the GCFT concern large and not small deviations, hence 
Gaussian distributions should not be expected.  For instance, these
distributions have finite support for $m=-1,0$, because the values of
$\zb_{l,m}$ are bounded, being a ratio between linear and quadratic
functions of the vorticity field. Furthermore, differently from
\cite{BoGaGa96}, the kurtosis of our distributions are manifestly
different from the Gaussian value $3$ (cf.\ Fig.\ref{betaden}) at
finite $\tau$'s, and we haven't observed any convergence to $3$ for
growing $\tau$. Even in the case of \cite{BoGaGa96}, in which a
Gaussian interpolation of the data is possible, a connection with the
central limit theorem is far from obvious.

{\bf 5.} Two recent papers, \cite{CL} and \cite{BiPiVu97},  considered
the verification of the EC and of the fluctuation relation in fluid
dynamics, suggested by Gallavotti \cite{Gal97}. The present work
strengthens these earlier findings, along with new results. In
\cite{CL}, one experimental fluctuating time series, representative of
a heat flux, is analyzed. The p.d.f.\ of this series is evidently
non-gaussian, but its odd part can be fitted by a linear relation, as
in our cases. The asymptotic value of the slope of the fitting lines
is not given.  In contrast, we were able to explore several cases and
to check how the asymptotic  $c_\sigma^{l,m}$ are reached.

The work of \cite{BiPiVu97}, instead, is mainly concerned with the
relation between ordinary shell models and  reversibly-damped
counterparts. The authors find differences in the  statistics produced
by the two versions of the shell models, as well as limitations on the
values which the parameters can take for one kind of equivalence to hold.
Such equivalence is found in the correct reproduction of the energy
cascade and of the multiscaling of the structure functions.  No
analysis of the GCFT is attempted there, at variance with our paper.
It is however interesting to compare our and their approach, as it
seems  that the dynamical equivalence between ``normal'' and
``reversibly  constrained'' dynamical systems representing fluid
turbulence is not always as complete as in the NS-GNS cases.

{\bf 6.} One final remark concerns the debate on the connection
between the GCFT and the so called Evans-Searles identity (ESI)
\cite{ES94,CohGal99}.  The ESI is a relation which concerns time
reversible dynamical systems, and the Liouville mesaure $\mu_L$ on the
phase space $\Omega$ of such systems. Hence it can also be applied to
our truncated GNS equations.  In particular, let $E_p \subset
\Omega$ be the subset of initial  conditions of trajectories along which
the phase space contraction after an
evolution of any length $T$ is  $e^{-p \zs_{l,m}^{GNS} T}$. Then, the  
ESI implies that \cite{CohGal99} 
\be 
 \log \mu_L(E_p) - \log \mu_L(E_{-p}) = p \langle\zs_{l,m}^{GNS} \rangle T ~.
\label{ESI}
\ee
This equation is formally similar to Eq.(\ref{finite})
re-written in terms of $\zs_{l,m}^{GNS}$, for $T=\zt$ 
and $C_\zs^{l,m}(\tau)=1$. 
Now, the fluctuation relation of the GCFT also has a slope of $1$, 
in the restricted domain of dynamical systems with dense attractors 
not considered here.
This led some to believe
that the ESI and the GCFT describe the same quantities, namely 
the fluctuations of the phase space contraction.
However, our results
explicitly show that the ESI cannot describe these fluctuations,
because we have
$c_\zs^{l,m}<1$. The ESI, instead, concerns the relative 
probability of independent ``trajectory histories'' emanating
from the Liouville distribution.

\section*{Acknowledgements}
We owe special gratitude to G.\ Gallavotti for suggesting this problem,
and for illuminating discussions throughout the time we devoted to this work.
We also thank G.\ Boffetta, F.\ Bonetto A.\ Celani and A.\ Vulpiani for 
stimulating and constructive criticism. We
thank the Institute for Cosmogeophysics of the CNR, Torino, for access to its
computer facilities. L.R.\ acknowledges support from GNFM-CNR (Italy) and 
through the EC contract ERBCHRXCT940460.

\bibliography{thermost}

\begin{thebibliography}{10}

\bibitem{Sp91}
H.~Spohn.
\newblock {\em Large scale dynamics of interacting particles}.
\newblock Springer, Berlin, 1991.

\bibitem{SJ}
Zhen-Su She and E.~Jackson.
\newblock Constrained {E}uler system for {N}avier-{S}tokes turbulence.
\newblock {\em Physical Review Letters}, 70(9):1255, 1993.

\bibitem{No84}
S.~Nos\`e.
\newblock A unified formulation of the constant temperature molecular dynamics
  methods.
\newblock {\em J. Chem. Phys.}, 81(1):511--519, 1984.

\bibitem{Ho85}
W.~G. Hoover.
\newblock Canonical dynamics: equilibrium phase-space distribution.
\newblock {\em Phys. Rev. A}, 31(3):1695--1697, 1985.

\bibitem{EM}
D.J. Evans and G.P. Morriss.
\newblock {\em Statistical mechanics of nonequilibrium liquids}.
\newblock Academic Press, London, 1990.

\bibitem{GalCoh95}
G.~Gallavotti and E.G.D. Cohen.
\newblock Dynamical ensembles in stationary states.
\newblock {\em Journal of Statistical Physics}, 80(5/6):931, 1995.

\bibitem{Gal95}
G.~Gallavotti.
\newblock Ergodicity, ensembles, irreversibility in {B}oltzmann and beyond.
\newblock {\em Journal of Statistical Physics}, 78:1571, 1995.

\bibitem{Gal97}
G.~Gallavotti.
\newblock Dynamical ensemble equivalence in fluid mechanics.
\newblock {\em Physica D}, 105:163, 1997.

\bibitem{RZ}
D.H. Rothman and S.~Zaleski.
\newblock {\em Lattice gas cellular automata}.
\newblock Cambridge University Press, Cambridge, 1997.

\bibitem{Ru78}
D.~Ruelle.
\newblock What are the measures describing turbulence?
\newblock {\em Prog. Theor. Phys. (supplement)}, 64:339, 1978.

\bibitem{CL}
S.~Ciliberto and C.~Laroche.
\newblock An experimental test of the {G}allavotti-{C}ohen fluctuation theorem.
\newblock {\em Journal de Physique IV}, 8(6):215, 1998.

\bibitem{BiPiVu97}
L.~Biferale, D.~Pierotti, and A.~Vulpiani.
\newblock Time-reversible dynamical systems for turbulence.
\newblock {\em Journal of Physics A}, 31:21, 1998.

\bibitem{BoGaGa96}
F.~Bonetto, G.~Gallavotti, and P.~Garrido.
\newblock Chaotic principle: an experimental test.
\newblock {\em Physica D}, 105:226, 1997.

\bibitem{LLP}
R.~Livi S.~Lepri and A.~Politi.
\newblock Energy transport in anharmonic lattices close to and far from
  equilibrium.
\newblock {\em Physica D}, 119:140, 1998.

\bibitem{BoChLe98}
F.~Bonetto, N.I. Chernov, and J.L. Lebowitz.
\newblock Global and local) fluctuations of phase space contraction in
  deterministic stationary non-equilibrium.
\newblock {\em http://xxx.lanl.gov/chao-dyn/9804020}, 1998.

\bibitem{ES94}
D.~J. Evans and D.~J. Searles.
\newblock Equilibrium microstates which generate second law violating steady
  states.
\newblock {\em Physical Review E}, 50:1645, 1994.

\bibitem{HoHoPo87}
B.L. Holian, W.G. Hoover, and H.A. Posch.
\newblock Resolution of {L}oschmidt's paradox: The origin of irreversible
  behavior in reversible atomistic dynamics.
\newblock {\em Physical Review Letters}, 59:10, 1987.

\bibitem{LlNiRoMo95}
J.~Lloyd, M.~Niemeyer, L.~Rondoni, and G.P. Morriss.
\newblock The nonequilibrium {L}orentz gas.
\newblock {\em CHAOS}, 5(3):536, 1995.

\bibitem{Ruelle96}
D.~Ruelle.
\newblock Positivity of entropy production in nonequilibrium statistical
  mechanics.
\newblock {\em Journal of Statistical Physics}, 85:1, 1996.

\bibitem{CoRo98}
E.G.D. Cohen and L.~Rondoni.
\newblock Note on phase space contraction and entropy production rate in
  thermostatted {H}amiltonian systems.
\newblock {\em CHAOS}, 1998.

\bibitem{RoCo}
L.~Rondoni and E.G.D. Cohen.
\newblock Orbital measures in nonequilibrium statistical mechanics: the
  {O}nsager relations.
\newblock {\em Nonlinearity}, 1998.

\bibitem{GaPRL}
G.~Gallavotti.
\newblock Extension of {O}nsager's reciprocity to large fields and the chaotic
  hypothesis.
\newblock {\em Physical Review Letters}, 78:4334, 1996.

\bibitem{Krai67}
Robert~H. Kraichnan.
\newblock Inertial ranges of two-dimensional turbulence.
\newblock {\em Physics of Fluids}, 10:1417--23, 1967.

\bibitem{CaHuQuZa88}
C.~Canuto, M.~Y. Hussaini, A.~Quarteroni, and T.~A. Zang.
\newblock {\em Spectral methods in Fluid dynamics}, volume~52 of {\em Springer
  series in computational Physics}.
\newblock Springer, Berlin, 1988.

\bibitem{KraMo80}
Robert~H. Kraichnan and David Montgomery.
\newblock Two-dimensional turbulence.
\newblock {\em Reports on Progress in Physics}, 43:547--619, 1980.

\bibitem{Leith84}
C.~E. Leith.
\newblock Minimum enstrophy vortices.
\newblock {\em Physics of Fluids}, 27(6):1388--1395, 1984.

\bibitem{MiWeCr92}
Jonathan Miller, Peter~B. Weichman, and M.~C. Cross.
\newblock Statistical mechanics, {E}uler's equation and {J}upiter's red spot.
\newblock {\em Physical Review A}, 45(4):2328--2359, 1992.

\bibitem{EvCoMo93}
D.J. Evans, E.G.D. Cohen, and G.P. Morriss.
\newblock Probability of second law violations in shearing steady flows.
\newblock {\em Physical Review Letters}, 78:434, 1993.

\bibitem{CohGal99}
E.~G.~D. Cohen and G.~Gallavotti.
\newblock Note on two theorems in statistical mechanics.
\newblock {\em http://xxx.lanl.gov/cond-mat/9903418}, 1999.

\end{thebibliography}
\bibliographystyle{unsrt}

\end{document}